# Structural and magnetic properties of irradiated SiC

Yutian Wang[1, 4], Xuliang Chen[2], Lin Li[1, 3], Artem Shalimov[1], Wei Tong[5], Slawomir Prucnal[1], Frans Munnik[1], Zhaorong Yang[2], Wolfgang Skorupa[1], Manfred Helm[1,4], Shengqiang Zhou[1]

1. Institute of Ion Beam Physics and Materials Research, Helmholtz-Zentrum Dresden-Rossendorf (HZDR), P.O. Box 510119, 01314 Dresden, Germany
2. Key Laboratory of Materials Physics, Institute of Solid State Physics, Chinese Academy of Sciences, Hefei 230031, People's Republic of China
3. Department of Physics and Electronics, School of Science, Beijing University of Chemical Technology, Beijing 100029, China
4. Technische Universität Dresden, 01062 Dresden, Germany
5. High Magnetic Field Laboratory, Hefei Institutes of Physical Science, Chinese Academy of Sciences, Hefei 230031, People's Republic of China

**Abstract:**

We present a comprehensive structural characterization of ferromagnetic SiC single crystals induced by Ne ion irradiation. The ferromagnetism has been confirmed by electron spin resonance and possible transition metal impurities can be excluded to be the origin of the observed ferromagnetism. Using X-ray diffraction and Rutherford backscattering/channeling spectroscopy, we estimate the damage to the crystallinity of SiC which mutually influences the ferromagnetism in SiC.

## I. Introduction:

Defect-induced ferromagnetism in materials which do not contain partially filled 3*d* or 4*f* electrons has recently entered the focus of condensed matter research [1]. It challenges the *mJ* paradigm for magnetism, where *m* refers the local moment and *J* stands for the interaction between the local moments. Experimentally, defect-induced ferromagnetism was observed in many materials, including graphite [2-5] and various oxides [6-12]. SiC single crystals are emerging as another candidate for this investigation and have been shown to be ferromagnetic after particle irradiation [13, 14] or after aluminum doping [15]. The magnetization in SiC is found to sensitively depend on the fluence of ion or neutron irradiation. Compared with other materials, SiC is commercially available at large scale with the microelectronic quality grade [16]. In this paper, we present a systematic structural investigation in correlation with the magnetic properties of SiC prepared by Ne ion irradiation. A possible Fe, Co or Ni contamination in SiC is excluded by particle induced X-ray emission (PIXE) and by Auger electron spectroscopy (AES) measurements. The appearance of ferromagnetism has been confirmed by electron spin resonance (ESR) spectroscopy. Measurements of X-ray diffraction (XRD) and Rutherford backscattering/channeling spectroscopy (RBS/C) reveal the damage to the crystallinity of SiC which can extinguish the ferromagnetism in SiC.

## II. Experimental methods:

The *6H*-SiC(0001) wafer purchased from KMT corporation (Hefei, China) is one-side polished and semi-insulating. The same wafer was cut into smaller pieces which were implanted with Ne ions at an energy of 140 keV. The ion fluences were $5\times10^{13}$, $1{\times}10^{14}$, $5{\times}10^{14}$, $1{\times}10^{15}$ cm$^{-2}$, and consequently the samples will be named as 5E13, 1E14, 5E14 and 1E15, respectively. Magnetometry was performed using a MPMS-XL magnetometer from Quantum Design. The ferromagnetic resonance was measured at 9.46 GHz by an electron paramagnetic resonance spectrometer (Bruker ELEXSYS E500). PIXE was performed using 3 MeV protons with a broad beam of 1 mm$^2$, while AES was done by a scanning Auger electron spectrometer Microlab 310F (Fisons Instruments). In order to evaluate the crystalline variation after ion irradiation, synchrotron radiation XRD was carried out at the Rossendorf beamline (BM20) at the ESRF with an X-ray wavelength of 0.1078 nm. As a complementary method, RBS/C with 1.7 MeV $He^+$ was used to quantitatively determine the ion-implantation-induced atomic disorder of the Si sublattice in 6H-SiC.

## III. Results and discussion:

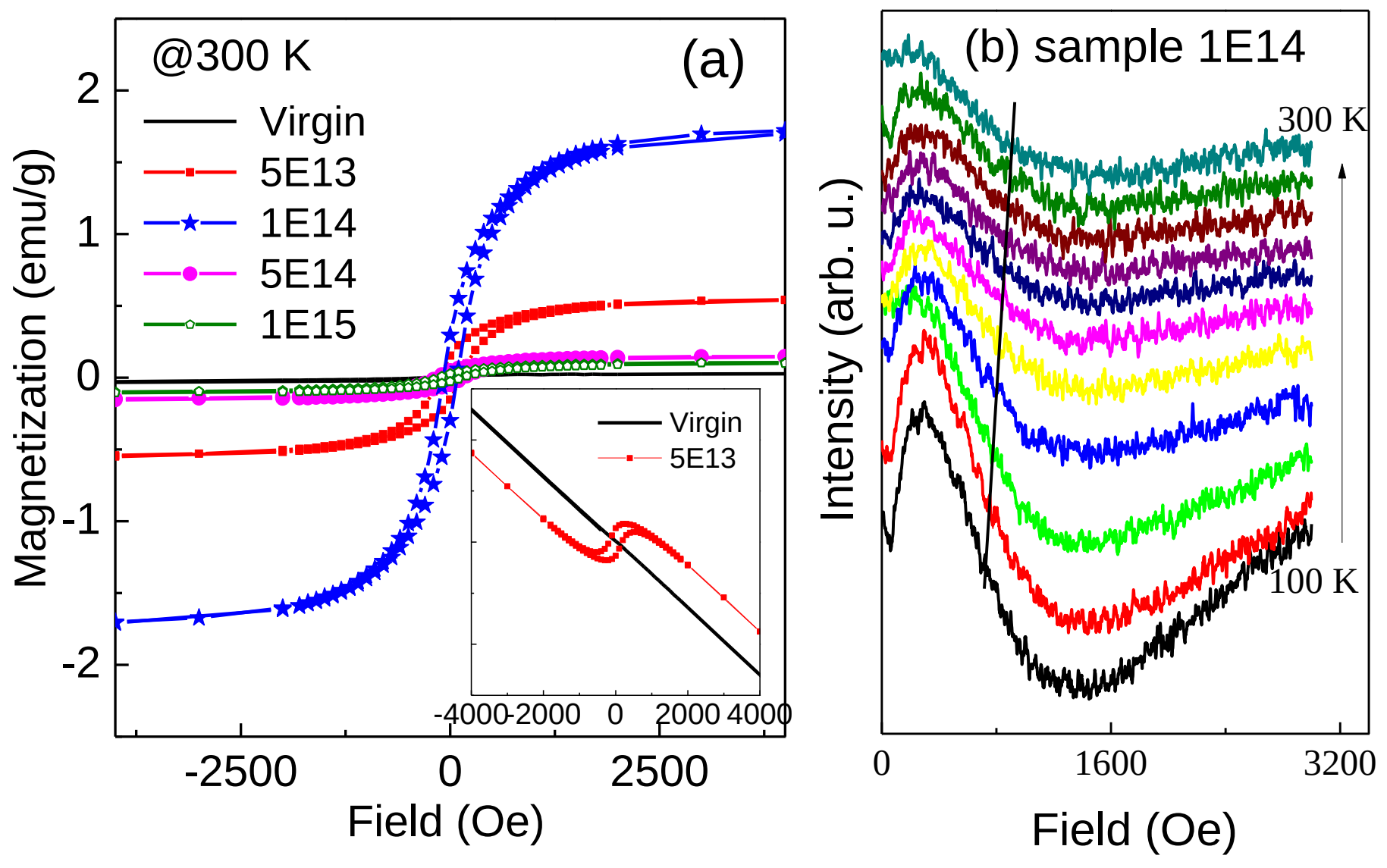

Fig. 1: (a) Magnetization vs. field measured at 300 K for all samples: the diamagnetic background from the substrate has been subtracted. The magnetization was calculated by assuming a thin layer of 460 nm thickness. The inset shows the magnetization measured for the virgin sample (black line) and sample 5E13 without subtracting the background. (b) ESR spectra of sample 1E14 at temperatures between 100 and 300 K: a broad resonance peak appears indicating ferromagnetic resonance. The black line indicates the resonance centers which slightly change with temperature.

### A. Magnetic properties

The magnetic properties of Ne irradiated samples at different temperatures have been reported in our previous paper [13]. The inset of Fig. 1(a) shows the magnetization at 300 K measured for sample 5E13 and the virgin SiC from the same wafer. The virgin SiC only shows diamagnetism, while sample 5E13 presents a ferromagnetic hysteresis. Fig. 1(a) shows the magnetization at 300 K for all samples after subtracting the diamagnetic background. The saturation magnetization shows a nonlinear dependence with increasing Ne fluence (defect concentration) and reaches its maximum for sample 1E14.

The appearance of the ferromagnetism in ion irradiated SiC is further confirmed by ESR. We have measured virgin SiC, sample 1E14 and one paramagnetic SiC in a broad field range. Only for sample 1E14, a broad ESR peak, and a shift of the center of the resonance at lower temperatures are clearly observed (Fig. 1b). Even at 300 K, the ESR signal shows a very large shift from the free-electron position (around 3300 Oe), showing that the sample is ferromagnetic and has $T_c$ above 300 K. The *g*-factor increases slightly with decreasing temperatures.

### B. Scrutinizing possible magnetic contaminations

For the investigation of defect-induced ferromagnetism, the crucial issue is to carefully examine if the ferromagnetic contamination is actually responsible for the observed ferromagnetism. The magnetization [Fig. 1(a)] shows a clear dependence on the ion fluence, however the magnetic contaminations are expected to be random or to increase with ion fluence if they are due to ion implantation. We further applied PIXE and AES to scrutinize Fe/Co/Ni elements in our studied SiC specimens. Figure 2 shows the PIXE spectra for virgin SiC and 5E13. In the spectrum, the narrow peak is from Si K-line X-ray emission. The broad peak is due to the Bremsstrahlung background. No transition metal impurities were detected within the detection limit of around 1 µg/g.

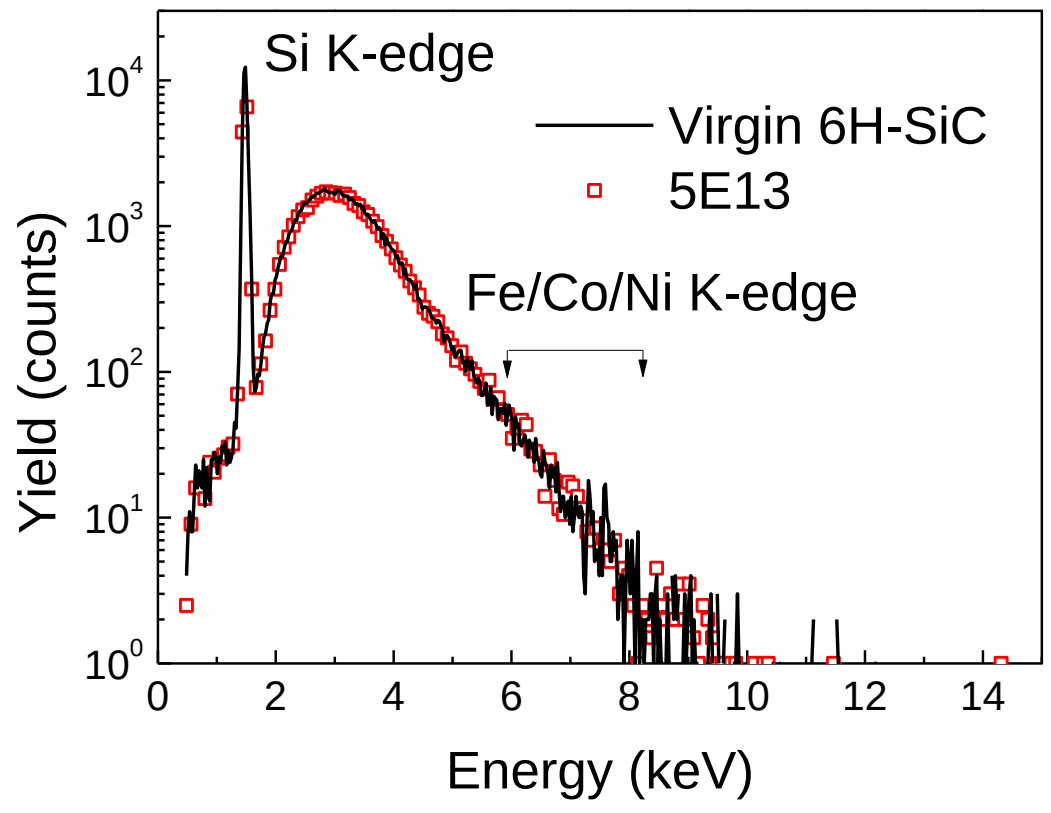

Fig. 2 PIXE spectrum for the virgin SiC wafer and a ferromagnetic sample. Within the detection limit, we do not observe any Fe, Co or Ni contamination.

To exclude contamination introduced during implantation and handling of the samples, we applied the surface sensitive AES method to the ferromagnetic pieces. For most elements, the detection limit of AES is in the range of 0.01-0.1 at%, which is around $10^{12}$-$10^{13}$ atom/cm$^2$ by assuming a detection depth of 2 nm. We measured both the front and back sides for virgin and ferromagnetic SiC after Ne irradiation and did not observe any signal from Fe/Co/Ni within the detection limit (not shown).

**C. Structural properties**

In order to correlate magnetization with structure, we performed XRD and RBS/C to check the crystalline deformation and defects.

Figure 3 depicts the $\theta$–$2\theta$ scans at SiC(000 12) recorded for Ne ion irradiated 6H-SiC. A virgin sample is also shown for comparison. All curves show a main sharp diffraction peak at the diffraction angle $\theta = \theta_{Bragg}$ coming from the sample volum below the irradiated part. The diffraction peaks at the low angle side of the main Bragg peak, ($\theta<\theta_{Bragg}$) are the characteristics of an expansion gradient of the lattice along SiC[0001]. The elastic strain in the near surface region ($\Delta d/d$) can be deduced from the position of this satellite peak. Except for the largest fluence applied here (1E15), a fringe pattern is clearly visible. This feature is due to the presence of the non-homogeneous lattice expansion along the depth and arises from interferences between x-ray beams diffracted from regions with the same lattice spacing. The maximum value of the normal strain $(\Delta d/d)_M$ is estimated to be roughly proportional to the fluence. Furthermore, we also performed the corresponding reciprocal spacing mapping (RSM) around SiC(000 12) (see Figure 4). Only a vertical streak coming from the irradiated region of the crystals is observed. No broadening in the horizontal direction (equivalent to a rocking curve) is measured as compared with that obtained from virgin crystals. These observations indicate that the defects are randomly distributed point defects or small defect clusters [18].

For the sample irradiated with the largest fluence of 1E15 $cm^{-2}$, the absence of the satellite peak shows that the near-surface region of this sample is heavily damaged, *i.e.* no more crystalline. Such a signal is consistent with a strongly defective crystalline lattice characterized by the presence of extended defects: amorphous clusters or layer [18]. The conclusions from XRD observation are fully consistent with the observation by positron annihilation spectroscopy, which reveals the clustering of vacancies [13].

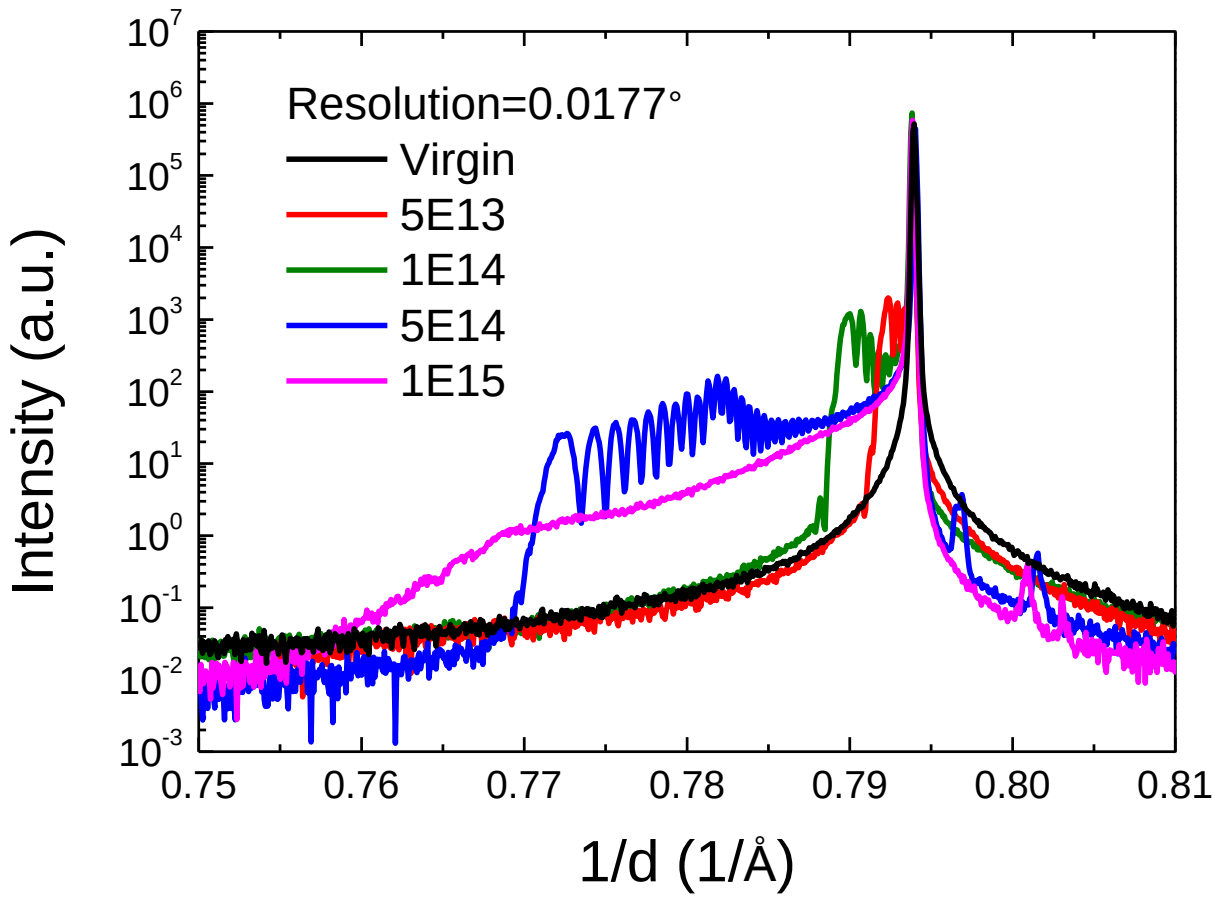

Fig. 3 XRD 2θ/θ scans of the virgin and Ne irradiated 6H-SiC recorded in the vicinity of the (000 12) reflection.

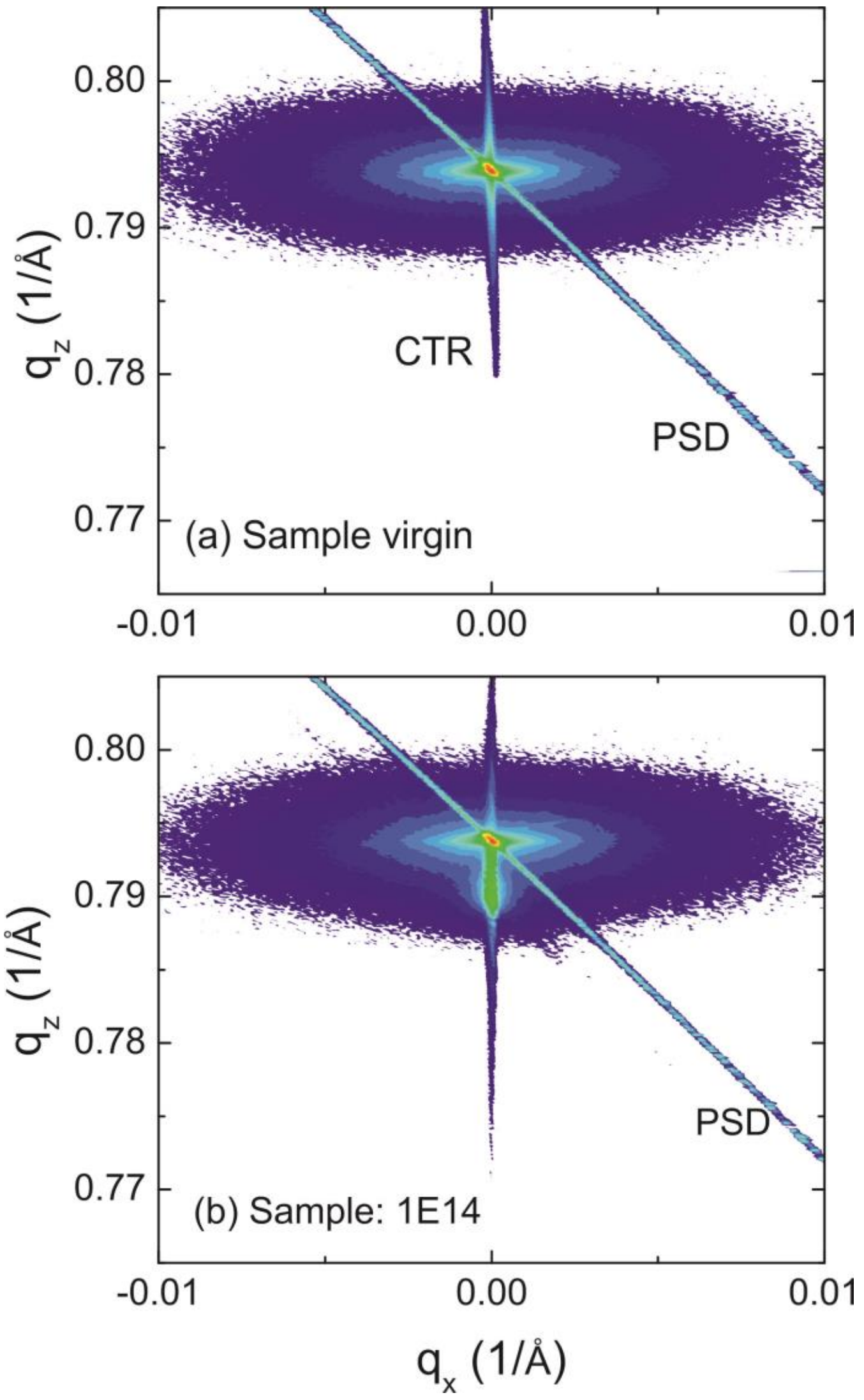

Fig. 4 Reciprocal space mapping of SiC(000 12). There is no indication of broadening along $q_x$. The streak labeled CTR is the crystal truncation rod. The streak labeled PSD is due to the transmittance function of the position sensitive detector.

The RBS/C spectra along the SiC[0001] axis are shown in Fig. 5 for virgin and for Ne irradiated samples. The random spectrum is assumed equivalent to a completely amorphized SiC, while the virgin spectrum corresponds to an essentially damage-free crystal. The minimum yield ($\chi_{min}$), the ratio of the channeling spectrum to the random spectrum, is 2.3% for the virgin SiC crystals at the surface region. For sample 5E13 (not shown), the channeling spectrum is only slightly higher than the virgin sample. A peak in the damage depth around 150 nm is becoming visible for fluences of $5\times10^{14}$ and $1\times10^{15}$ /cm$^2$. Note that for sample 1E15 $\chi_{min}$ is 93%, indicating the almost full amorphization at the damage peak. This is in agreement with the threshold displacement per atom (DPA) values for amorphization reported in literature [19]. DPA presents the irradiation effect considering both ion fluence and energy [20].

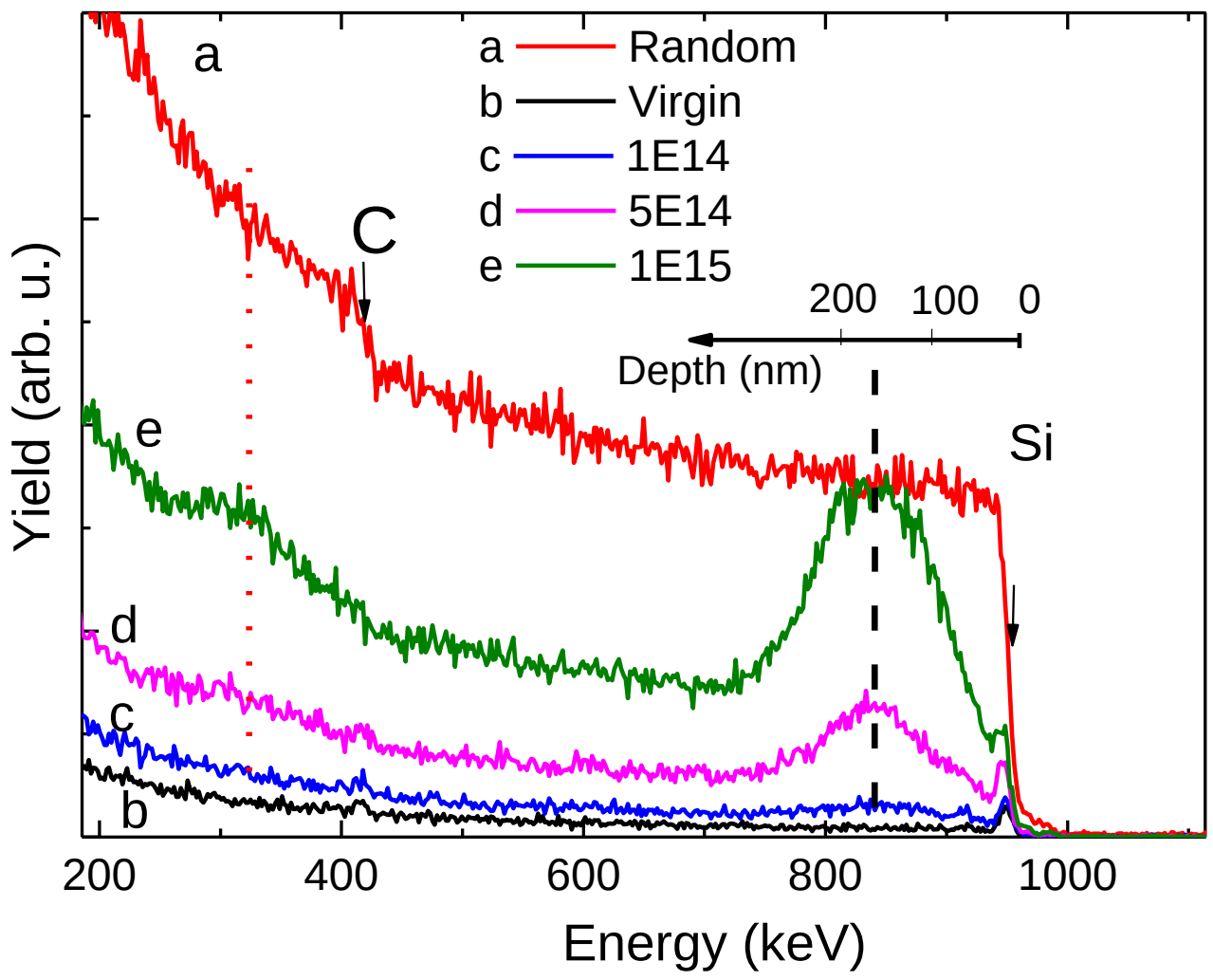

Fig. 5: RBS/C spectra for 6H-SiC implanted with Ne ions. A random spectrum and a channeling spectrum from a virgin sample are also included for comparison. The dashed (dotted) line indicates the position of the damage peaks for Si (carbon). A rough depth scale is given.

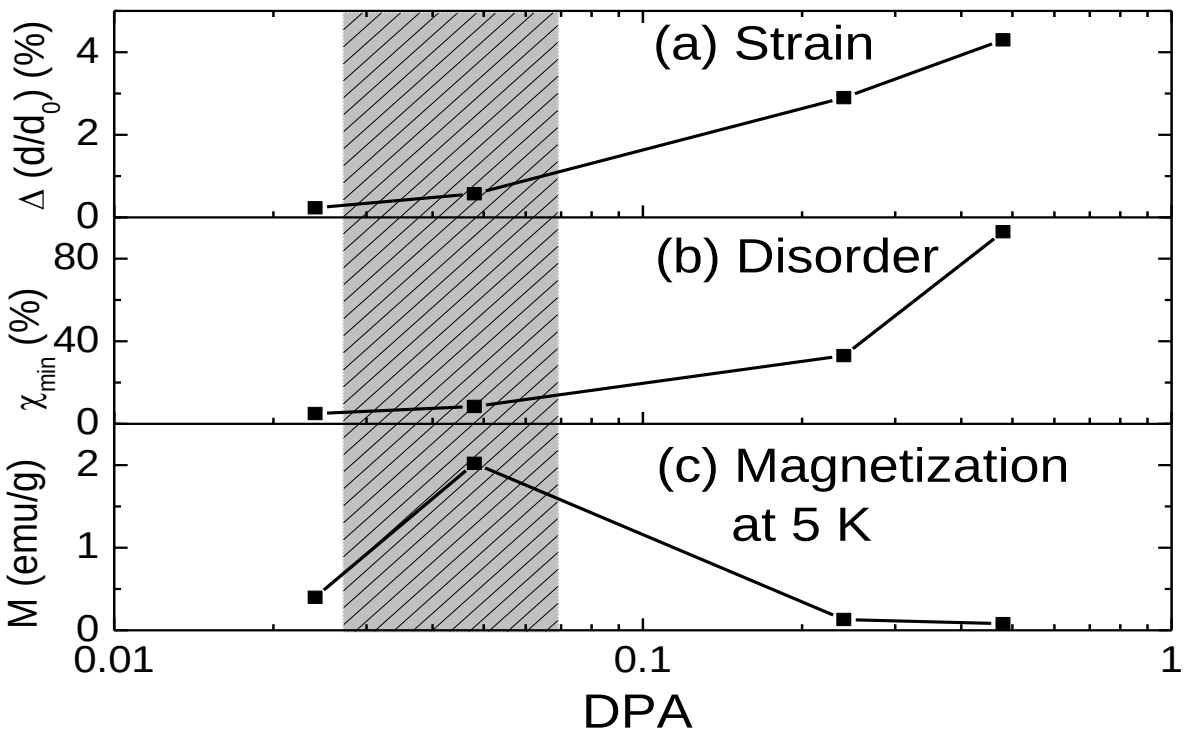

Fig. 6: Evolution of the saturation magnetization, the Si-sublattice disorder and the maximum strain with DPA. The gray area indicates the DPA range for optimizing the ferromagnetism.

## IV. Summary

Combining both structural and magnetic properties of ion irradiated SiC, we are able to conclude a mutual relation between defect concentration and magnetization. As shown in Fig. 6, there is a narrow DPA window, within which the sample is ferromagnetic. The initial introduction of structural disorder (mainly point defects) leads to pronounced magnetization. Further increasing disorder induces agglomeration of point defects and finally amorphizes the host SiC crystal, consequently the ferromagnetism nearly drops to zero.

Acknowledgement

The work was financially supported by the Helmholtz Association (VH-NG-713). Y. Wang thanks the China Scholarship Council for supporting his stay at HZDR.

Fig. captions

Fig. 1: (a) Magnetization vs. field measured at 300 K for all samples: the diamagnetic background from the substrate has been subtracted. The magnetization was calculated by assuming a thin layer of 460 nm thickness. The inset shows the magnetization measured for the virgin sample (black line) and sample 5E13 without subtracting the background. (b) ESR spectra of sample 1E14 at temperatures between 100 and 300 K: a broad resonance peak appears indicating ferromagnetic resonance. The black line indicates the resonance centers which slightly change with temperature.

Fig. 2 PIXE spectrum for the virgin SiC wafer and a ferromagnetic sample. Within the detection limit, we do not observe any Fe, Co or Ni contamination.

Fig. 3 XRD 2θ/θ scans of the virgin and Ne irradiated 6H-SiC recorded in the vicinity of the (000 12) reflection.

Fig. 4 Reciprocal space mapping of SiC(000 12). There is no indication of broadening along $q_x$. The streak labeled CTR is the crystal truncation rod. The streak labeled PSD is due to the transmittance function of the position sensitive detector.

Fig. 5: RBS/C spectra for 6H-SiC implanted with Ne ions. A random spectrum and a channeling spectrum from a virgin sample are also included for comparison. The dashed (dotted) line indicates the position of the damage peaks for Si (carbon). A rough depth scale is given.

Fig. 6: Evolution of the saturation magnetization, the Si-sublattice disorder and the maximum strain with DPA. The gray area indicates the DPA range for optimizing the ferromagnetism.